\documentclass{article}
\usepackage[nonatbib,preprint]{neurips_2020}
\usepackage[utf8]{inputenc} 
\usepackage[T1]{fontenc}    
\usepackage{hyperref}       
\usepackage{url}            
\usepackage{booktabs}       
\usepackage{amsfonts}       
\usepackage{nicefrac}       
\usepackage{microtype}      
\usepackage{wrapfig}
\usepackage{graphicx}
\usepackage{subcaption}
\usepackage{amsmath}
\usepackage{amssymb}
\usepackage[font=small]{caption}

\title{Learned Indexes for a Google-scale Disk-based Database}

\author{
  Hussam Abu-Libdeh\\
  Google
  \And
  Deniz Alt{\i}nb\"{u}ken\\
  Google
  \And
  Alex Beutel\\
  Google
  \And
  Ed H. Chi\\
  Google
  \And
  Lyric Doshi\\
  Google
  \And
  Tim Kraska\\
  MIT, Google
  \And
  Xiaozhou (Steve) Li\\
  Google
  \And
  Andy Ly\\
  Google
  \And
  Christopher Olston\\
  Google\footnotemark
}

\begin{document}
\nolinenumbers

\maketitle

\renewcommand{\thefootnote}{\fnsymbol{footnote}}
\footnotetext[1]{Work done while at Google, although the author currently does not work at Google.}
\renewcommand{\thefootnote}{\arabic{footnote}}

\vspace{-3mm}
\begin{abstract}
    There is great excitement about learned index structures, but understandable skepticism about the practicality of a new method uprooting decades of research on B-Trees. In this paper, we work to remove some of that uncertainty by demonstrating how a learned index can be integrated in a distributed, disk-based database system: Google's Bigtable. We detail several design decisions we made to integrate learned indexes in Bigtable. Our results show that integrating learned index significantly improves the end-to-end read latency and throughput for Bigtable.
\end{abstract}
\vspace{-4mm}

\section{Introduction}
\vspace{-2mm}
There has been a lot of excitement about the potential benefits of using machine learning methods in databases, but understandable skepticism about the practicality of these new methods for real applications. Two years ago learned index structures \cite{kraska2018case} were proposed as a novel technique to replace B-Trees and similar range index structures. Learned index structures use a learned CDF-model instead of the traditional tree-structures to achieve significantly faster lookup times and index space savings. While there has since been a flurry of follow up research, the impact of the work on real-world systems has been under question. The most common points of criticism include, that (1) the original learned index didn't support inserts,  (2) wasn't designed for disk-based systems, where indexes matter the most, and, more generally, (3) the index size or speed improvement do not really matter in practice as it only makes up a small fraction of the entire query execution time. While support for inserts has already been addressed by follow-up work \cite{alex,pgm,radix-spline}, the question of whether learned indexes can have an impact on real, disk-based systems remained unaddressed.

In this paper, we present initial results on integrating learned index structures into Google's Bigtable \cite{bigtable2006,bigtable2008} and show that they can significantly improve the read latency and throughput, especially at the tail. This is a surprising result as Bigtable is a distributed and disk-based system that has been optimized for high-throughput and low-latency reads and writes. Requests usually take tens to hundreds of milliseconds rather than nanoseconds. How can a learned index structure, which improves request times by nanoseconds in \cite{kraska2018case}, have a positive impact on the overall latency of such a system? While the faster lookup time has little impact, we see big performance gains from the second-order effects of using a learned index, such as reducing the index size, requiring fewer index seeks, reading fewer index blocks, and simplifying data block prefetching.

Bigtable uses a traditional B-Tree to determine which key is stored in which block. Hash maps are not an option in Bigtable as the system has to support range scans. Like many database systems, the B-Tree itself is also stored in the same block structure and can consist of several blocks. As a result, a request for a key might involve several block reads; usually 1-2 to read the index blocks and then 1 more to read the block with the data. Caching can significantly reduce the number of blocks to fetch from the distributed file system, especially the index blocks. However, in practice (1) Bigtable instances can become really big, so big that the index might not fit into main memory, (2) machines might host several Bigtable instances, reducing the amount of available memory, (3) failures and load-balancing might clear the cache, and (4) some Bigtable instances are both large and rarely used. All these factors create pressure on the cache and can cause index blocks to be fetched again. Reducing the index size by orders of magnitude can have an overall positive impact, especially on the tail latency. With a smaller index, fewer blocks have to be fetched even with a cold cache. The smaller index size also reduces the overall resource consumption and improves throughput.

This paper explores the opportunity of learned indexes to reduce the index size for distributed database systems in general, and Bigtable in particular. In order to integrate learned index structures into Bigtable, we had to significantly extend the original learned index design \cite{kraska2018case} for disk-based systems with block identifiers.
The original learned index assumed that all data is stored in a large, continuous array in memory, thus avoiding the non-trivial space cost of storing any pointers and allowing for cheap local searches following mispredictions of data locations. Unfortunately, this is not possible in Bigtable, which uses block-based files to store data and does not allow laying out data in one continuous array. We overcome this issue by using the model to \emph{define} which records are stored in which block, as opposed to learning which records are in which block. Therefore, model predictions are always correct. We also create a novel compressible mapping table for the block pointers.

In summary, we make the following contributions:
\vspace{-2mm}
\begin{enumerate}
    \item We present the first design of learned indexes for a distributed, disk-based database.
    \item We offer a new learned index model design for string keys with monotonicity guarantees.
    \item We present the evaluation of our method with uniform value sizes to understand the trade-offs in read performance. We demonstrate using learned indexes improves latency $36\%$ for point lookups and $22\%$ for scans on average. Similarly, using learned indexes increases throughput by $55\%$ for point lookups and $28\%$ for scans on average.
\end{enumerate}
\vspace{-3mm}

\section{Related Work}
\vspace{-2mm}
Learned index structures were first introduced in~\cite{kraska2018case}. Several works on learned index structures followed~\cite{pgm,fiting_tree,alex,radix-spline,lis_func_interp,learnedspatial,learnedspatial2,flood,tsunami,sosdb}.
For example, PGM index~\cite{pgm}, FITing tree~\cite{fiting_tree}, RadixSpline~\cite{radix-spline} and ALEX~\cite{alex} all investigate how to efficiently support inserts, provide better worst-case guarantees, and/or  faster index construction.
Lisa \cite{lisa} extends the idea of learned index structures for DNA Sequence search,   \cite{learnedspatial,learnedspatial2} for spatial applications, and Flood~\cite{flood} and Tsunami~\cite{tsunami} for multi-dimensional data. More recently, the SOSD benchmark \cite{sosdb} demonstrated that learned index structures outperform their traditional counterparts on a wide range of datasets in size and lookup latency.
However, none of these algorithms are designed for disk-based or distributed systems.

Most closely related to this work is BOURBON \cite{BOURBON}, which shows how to integrate learned index structure for an LSM-based Key-Value Store. A log-structure merge (LSM)-tree is a highly write optimized data structure for disk. However, their system is a research prototype - so it is still not answering the question if learned indexes would impact a real production system, nor are their techniques targeted toward distributed systems and read-heavy workloads.
\vspace{-3mm}
\section{Background System Design}
\vspace{-2mm}
\textbf{Bigtable} \cite{bigtable2008} is a distributed structured storage system that serves petabytes of data. Bigtable can support jobs that need to meet high throughput requirements as well as jobs that require low latency, using a server that manages the Bigtable cell, and many tablet servers that serve data.

Writes in Bigtable are handled directly by the tablet servers, where they are first committed to a log on disk, then written to the memtable. When the memtable size reaches a threshold, it is \emph{compacted} to disk as a read-only SSTable. Reads are also handled directly by the tablet servers either from the memtable or SSTables. During reads, SSTables are loaded into memory from a distributed file system and scanned to find the correct value.

\textbf{SSTables} are structured as a map of key-value pairs, sorted by keys of the form (row, column, timestamp). To support efficient operations on very large data sets, SSTable implementation partitions data into data blocks and keeps an index of which keys reside in which data blocks. Reads can then use the SSTable index to find the correct data blocks for a given key.

The SSTable index is stored in index blocks as an ordered mapping from the last key in a data block to the associated index entry for that data block. The index entry contains information about the location and size of the block. Using these, the data block can be efficiently loaded into memory. The index block is intended to stay in memory, meaning reads, especially continuous reads, can be fulfilled by querying the index block and only making directed disk reads for the requested data.

The SSTable index is created while SSTables are constructed. During construction, once a data block has been filled up, an appropriate entry is added to the index. The data block is then written out and evicted from memory. To finalize the SSTable, the index is written out as well.

With growing SSTable sizes, the size of the index block can also grow. To improve index efficiency, SSTables use two-level indexes, comprising a level-0 index and a level-1 index. The level-0 index spans a single index block and always resides in memory, pointing to level-1 index blocks. The level-1 index is a sequence of level-1 index blocks, which point to data blocks. The level-1 index blocks are loaded into memory when a data block that they point to is accessed.

\vspace{-3mm}
\section{Learned Indexes for Disk-based Databases}
\vspace{-2mm}
\paragraph{Design}
As described above, an SSTable $\mathcal{D}$ consists of key, value pairs $r = \langle k, v \rangle$.  These records are sorted by the key $k$ and stored into blocks. A traditional look-up index searches through the index data structure to identify which block contains the record for a given key. The learned index design in \cite{kraska2018case} proposed the index tries to predict the record's location directly by a given key $k$. If the predicted location was incorrect, the system needed to search nearby (e.g. with exponential search) to find the requested record.  Unfortunately, this doesn't work well for Bigtable.  Predicting the exact location of a record is not useful since the system needs to read the entire block. As a result, we instead focus on the learned index predicting blocks not records.  Even then, if the predicted block is incorrect, searching nearby would require reading whole new blocks, which is much more costly.

Therefore, we take a new approach.  Let's assume we are training a learned index $f(\cdot)$ to take a key $k$ and predict a block number. Given a learned index $f$, we designate block boundaries during SSTable creation such that each record $\langle k, v \rangle$ is placed in predicted block $f(k)$. By writing the data to match the index, we guarantee that the learned index always predicts the correct block for a given key. Further, by using a monotonic $f$, we ensure that records are still sorted by key and we can still perform range scans.

\vspace{-2mm}
\paragraph{Learned Index Training}
How should we train the learned index to predict the block number before the block number is set?  Rather than train the learned index to predict the block number, we  train it to predict how many bytes into the data the record falls.  We can use this estimation of number of bytes to build approximately equal-sized blocks.

To make this precise, assume that $s(\langle k, v \rangle)$ is the number of bytes in record $\langle k, v \rangle$ and let $S(\langle k, v \rangle)$ denote the number of bytes preceding $k$ in the SSTable $\mathcal{D}$:
\begin{align}
    S(\langle k, v \rangle) = \sum_{\langle k', v' \rangle \in \mathcal{D} | k' < k} s(\langle k', v' \rangle)
\end{align}
We use $S(\langle k, v \rangle)$ as the supervision for training $f$, i.e. $\min_f \sum_{\langle k, v \rangle} (f(k) - S(\langle k, v \rangle))^2$.  Given a prediction of the number of bytes into the data a record lies, we can recover the predicted block number by $\lfloor f(k) / \tau \rfloor$ where $\tau$ is the desired average block size.
While model inaccuracy may result in a less uniform distribution of block sizes, the model will never predict the wrong block.

\vspace{-2mm}
\paragraph{Model Structure}
The results presented in this paper are based on a linear regression model, but any model can be used (e.g., RMIs \cite{kraska2018case}). The simplicity of the model reduces both storage and inference costs while providing the necessary monotonicity guarantees over string keys.

Since the keys are strings of varying lengths, we convert them into an integer for input to the linear regression.  In this conversion, we treat the leading characters of the string as the leading bits of the integer. Doing this naively would mean we could only take into account the first few characters in the key. Instead, we shift and re-base on a character by character basis, based on the range of characters observed at each character position. During training, we determine the minimum and maximum ASCII value of each character position over all keys. We then convert the keys into integers by choosing a different numerical base for each character position according to the range of values that position can take. To give an example, consider first how we encode numbers in base 10: $132 = 1 * 100 + 3 * 10 + 2$. Now, consider that the only keys are "aab", "bdd", and "bcb". We need 2 values to encode the range of the $0^{th}$ position, 4 values to encode the range of the $1^{st}$, and 3 values to encode the range of the $2^{nd}$. Therefore, "bcb" becomes $1 * (4 * 3) + 2 * (3) + 0$. These converted keys become the input $x$ for an Ordinary Least Squares (OLS) model matching those $x$ to $S(\langle k, v \rangle)$. We can compute the parameters in closed form after a single pass over the data.

\vspace{-2mm}
\paragraph{Block Locations}
As the learned index predicts block numbers instead of disk locations, we need a way to map a predicted block number to a disk location.  For this, we store the location of each block location in an array $B$ such that $B[ \lfloor f(k) /\tau \rfloor]$ outputs the location of the block on disk.  Because the blocks are stored contiguously, $B[ \lfloor f(k) /\tau \rfloor + 1] - B[ \lfloor f(k) /\tau \rfloor]$ makes the block size available.  For backwards compatibility with traditional indexes in Bigtable, we also store additional metadata (e.g. checksums) for each block.  While this design may not seem more efficient than the last layer of a traditional index, it saves significant space in two regards: (1) none of the keys need to be stored (2) the disk locations can be compressed as they are approximately uniformly spaced\footnote{In the experiments below we don't use a compressed form of this map.}.

\vspace{-3mm}
\section{Evaluation and Conclusion}
\vspace{-2mm}
We evaluate how using a learned index affects Bigtable read performance in comparison to using a two-level index (Bigtable's default). In our preliminary results, we measure point lookups and scans from 5 tablet servers, each with 4G RAM. For point lookups, 15 clients maintain a buffer of 100 asynchronous single row read requests each, sending new requests as they receive replies. For scans, every operation scans 100 consecutive rows of size 1KB. During reads, tablet servers prefetch index and data blocks as needed. The SSTables are constructed with random integer keys (encoded as strings). Each row key is associated with data of size 1KB, and a data block size of 32KB.

We ran experiments with tables of different sizes (16 to 512 million rows) and the size of the table did not impact the read performance. The results we are presenting are for a table size of 256 million rows.

\begin{figure}[!t]
\begin{subfigure}{.49\textwidth}
\minipage{0.5\textwidth}
  \centering
  \includegraphics[width=48pt]{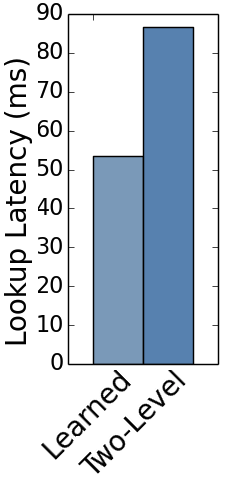}
  \centering
  \hspace{-0.5em}
  \includegraphics[width=48pt]{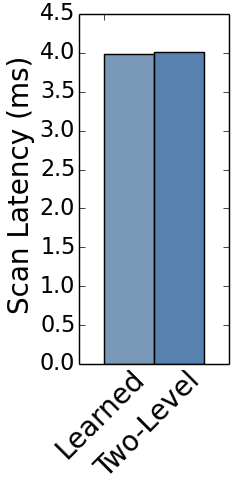}
  \caption*{(a) p99 Latency}
\endminipage
\minipage{0.5\textwidth}
  \centering
  \includegraphics[width=47pt]{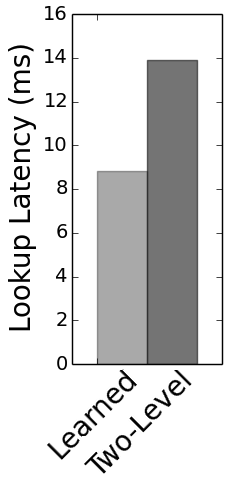}
  \centering
  \includegraphics[width=48pt]{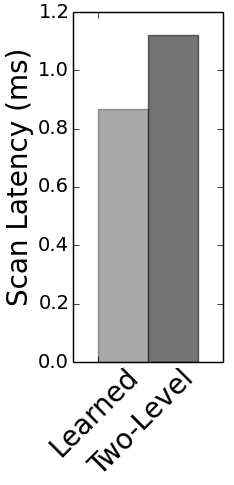}
  \caption*{(b) Mean Latency}
\endminipage
\end{subfigure}
\begin{subfigure}{.5\textwidth}
\minipage{0.5\textwidth}
  \centering
  \includegraphics[width=48pt]{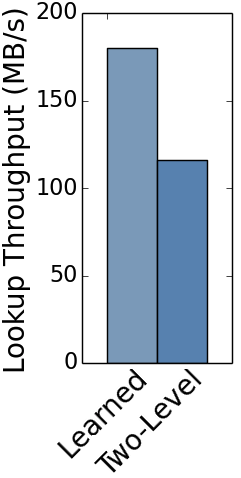}
  \centering
  \includegraphics[width=48pt]{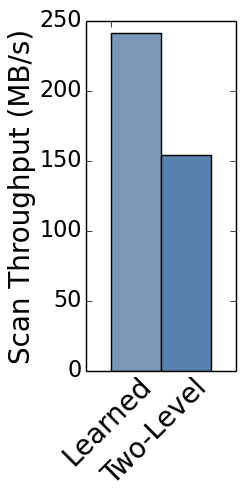}
  \caption*{(c) p99 Throughput}
\endminipage
\minipage{0.5\textwidth}
  \centering
  \includegraphics[width=48pt]{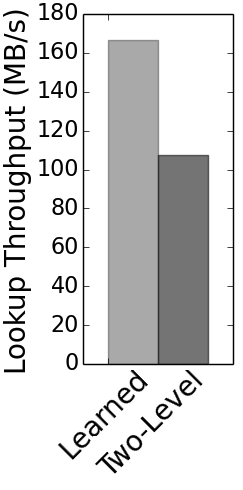}
  \centering
  \includegraphics[width=48pt]{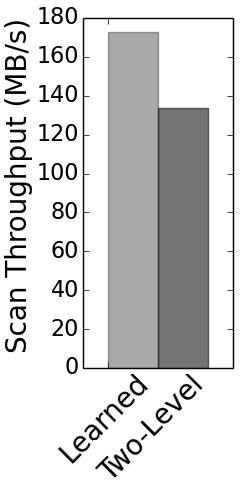}
  \caption*{(d) Mean Throughput}
\endminipage
\end{subfigure}
\caption{Learned index demonstrates performance improvements over the two-level index for lookups and scans. (a), (b): Latency reduced. (c), (d) Throughput improved. Scan numbers are normalized as per 1KB row.}
\label{fig:evaluation}
\vspace{-6mm}
\end{figure}

Figure \ref{fig:evaluation}(a) shows that the learned index reduces the read latency by $38\%$ for point lookups while maintaining it for scans in the 99th percentile. The mean latency was also reduced $36\%$ for point lookups and $22\%$ for scans as shown in Figure \ref{fig:evaluation}(b). Learned index eliminates the need for disk accesses for fetching index blocks, which improves the read latency for databases with larger indexes.

The learned index also significantly increases the throughput and indirectly, reduces CPU resource usage. Throughput increased $54\%$ for point lookups and $56\%$ for scans (see  Figure~\ref{fig:evaluation}(c)) in the 99th percentile. Comparably, the mean throughput increased by $55\%$ for point lookups and by $28\%$ for scans (see Figure~\ref{fig:evaluation}(d)). The reason for this is more subtle: Like many disk-based systems, Bigtable heavily compresses the data. Reducing the number of blocks accessed reduces the amount of work needed to decompress the data in addition to other secondary effects.

Early results from our ongoing experiments with different key distributions, workloads, and value sizes show similar performance improvements from using a learned index (see also \cite{sosdb}). Furthermore, using the learned index does not affect the write performance because the model training is decoupled from the write path in Bigtable. The learned index model is trained in the background during SSTable creation after memtables fill up.

In summary, our evaluation shows that learned indexes can improve real-world systems because of the cascading benefits of their significantly smaller size and simple usage.

\small
\bibliographystyle{abbrv}
\bibliography{bigtable_learned_index}
\end{document}